\documentclass[12pt,a4paper]{article}

\usepackage{amsthm,amsfonts,amssymb,latexsym,  graphics}

\usepackage{epsfig, euscript}

\title{Bell as the Copernicus of Probability}
\author{Andrei Khrennikov\\
International Center for Mathematical Modeling\\
in Physics and Cognitive Sciences \\
Linnaeus University, V\"{a}xj\"{o}-Kalmar, Sweden}

\begin{document}

\maketitle

\abstract{Our aim is to emphasize the role of mathematical models in physics, especially models of geometry and probability.
We briefly compare developments of  geometry  and probability by pointing to similarities and differences:
from Euclid to Lobachevsky and from Kolmogorov to Bell. In probability Bell played the same role 
as Lobachevsky in geometry. In fact,  violation of Bell's inequality implies the impossibility to apply 
the classical probability model of Kolmogorov (1933) to quantum phenomena. Thus 
quantum probabilistic model (based on Born's rule) is an example of non-Kolmogorovian model of probability, similarly 
to the Lobachevskian model -- the first example of non-Euclidean model of geometry. We also discuss coupling of the classical 
probabilistic model with classical (Boolean) logic. The Kolmogorov model of probability is based on the set-theoretic presentation 
of the Boolean logic. In this framework violation of Bell's inequality implies the impossibility 
to use the Boolean structure of events for quantum phenomena; instead of it,  events  have to be 
represented by linear subspaces. This is  the ``probability model'' interpretation  
of violation of Bell's inequality. We also criticize the standard interpretation -- an attempt to add to rigorous 
mathematical probability models additional elements such as (non)locality and (un)realism.  Finally, we compare embeddings of non-Euclidean
geometries into the Euclidean space with embeddings of the non-Kolmogorovian probabilities (in particular, quantum probability)
into the Kolmogorov probability space. As an example, we consider the CHSH-test.}

\section{Introduction}

\setcounter{equation}{0}

The argument which will be presented in this paper, namely, that violations of Bell type inequalities \cite{B0}, \cite{B} 
signal us that the classical model 
of probability \cite{K} (Kolmogorov, 1933) is inapplicable to quantum phenomena,   was already discussed in very detail 
by many authors, see, e.g., \cite{1}--\cite{1X}. Here I try to formulate this argument so clearly as possible. 
I also argue that any attempt to assign to these violations some additional value
e.g., to philosophize about (non)locality and (un)reality, is not consistent with the {\it mathematical model} approach  
to description of natural (and mental) phenomena (see \cite{AKHR} for a discussion).

The modern physics is based on creation of mathematical models describing natural phenomena. 
We emphasize that a model has not be identified with reality.  Each model has boundaries of its application.
Thus the evolution of physics can be considered as {\it a chain of creation of mathematical models, finding boundaries of their
applicability, and creation of new   models.}  Personally I think that any attempt to proceed without keeping to 
this mathematical modeling ideology leads to philosophizing which only shadows the structure of theory (although it may be generates 
the feeling of understanding).

From this viewpoint the Bell argument \cite{B0}, \cite{B} led to the recognition that the classical Kolmogorov model of probability 
\cite{K} which served 
so well for classical statistical physics has to be rejected and one has to use the quantum model of probability.  And (!) nothing more.   

Now we move to the issue of similarities and differences in the evolutions of geometry and probability and their impact to physics.
We recall that {\it Clifford called Lobachevsky the Copernicus of Geometry due to the revolutionary character of his work.} Lobachevsky 
``discovered'' the first non-Euclidean model which nowadays is  known as the Lobachevskian or hyperbolic geometry \cite{LB1}. We also point out
that he discussed possible experiments in astrophysics to check whether the geometry of Universe is Euclidean or Lobachevskian.\footnote{
In 1826 he noted that it is possible to find experimentally the deviation from $\pi $
of the sum of the angles of cosmic triangles of great size; in a later work  
he moved to the opposite scale and suggested that his geometry might find application in the intimate sphere of molecular attractions, see, e.g., 
\cite{LB2}.}

\begin{figure}[htpb]
\centering
\begin{minipage}[b]{0.45\linewidth}
\includegraphics[scale=0.23]{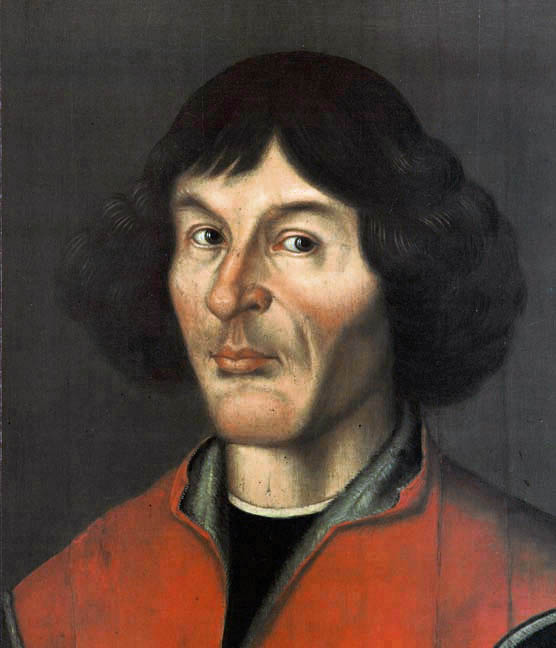}
\caption{Nicolaus Copernicus}
\label{fig:minipage1}
\end{minipage}
\quad
\begin{minipage}[b]{0.45\linewidth}
\includegraphics[scale=0.45]{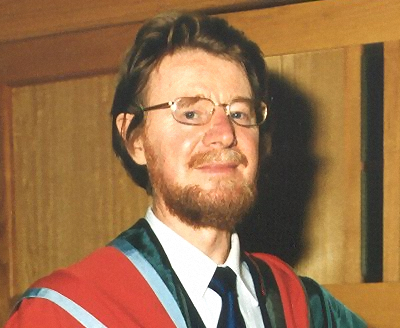}
\caption{John Bell}
\label{fig:minipage2}
\end{minipage}
\end{figure}

I would call {\it Bell the Copernicus of Probability}. In fact,  violation of Bell's inequality implies the impossibility to apply 
the classical probability model of Kolmogorov (1933) to quantum phenomena. Thus 
the quantum probability model (based on Born's rule) gives us an example of non-Kolmogorovian model of probability, similarly 
to the Lobachevskian model (an example of non-Euclidean model). However, there are not only similarities, but
 also differences. Physicists sufficiently quickly 
understood that a variety of models of geometry can be fruitfully explored. For example, the Lobachevskian geometry 
is used in special relativity. And the combination of the efforts of mathematicians (first of all, Riemann and Hilbert) and physicists 
(first of all, Einstein) led to the wide use of non-Euclidean geometries in general relativity. At the same time physicists 
have no (or very little) interest to the interpretation of quantum mechanics (QM) as a theory based on a special
 non-Kolmogorovian model of probability.

We shall also discuss coupling of the classical 
probabilistic model with classical (Boolean) logic. The Kolmogorov model of probability is based on the set-theoretic presentation 
of the Boolean logic, the crucial point is that events are represented by subsets of a ``set of elementary events'' $\Omega$ and operations 
for events are classical Boolean operations. In this framework violations of Bell-type inequalities can be interpreted simply as a signal 
that one cannot use this set-theoretic representation of events for quantum phenomena, instead of it, one has to represent events (detections
for quantum systems) by linear subspaces (orthogonal projectors) in a complex Hilbert space.       

All previous attempts, see especially  \cite{KB1}, \cite{book2}, to communicate with the physical community about Kolmogorovness/non-Kolmogorovness issue
of Bell's argument\footnote{One can also say ``Boolean/non-Boolean''. It is a good place to recall that J. Boole by himself 
analyzed a possibility that probabilities for some group of random observations cannot be embedded in the Boolean algebra \cite{Boole}. 
He even investigated the standard Bell's situation. There are given three dichotomous random observables. It is possible to perform 
their pairwise measurements and the corresponding joint probability distributions are given. The natural question arises: Is it always
possible to define the joint probability distribution for this triple of observables? Boole presented the inequality which is nowadays 
known as Bell's inequality as a necessary condition for the existence of this probability. Later this problem, of the existence of the 
joint probability distribution, was studied in very details by Vorob'jev \cite{vor}, for any group of observables taking the finite number of values.}    
 were not successful. One of the roots of such miscommunication is that majority of physicists 
are poorly educated in probability theory (it seems that students in physics  do not have a mathematically rigorous course 
in probability theory, neither in Europe nor in USA). Therefore we start the paper with a brief presentation of the modern axiomatic 
approach to probability theory \cite{K}.

Since classical probability is fundamentally based on classical logic, rejection of the Kolmogorov model of probability automatically implies rejection 
of the Boolean model of logic. Thus Bell's ``no-go theorem'' tells us that for analysis of results of quantum observations it is impossible to use 
the Boolean laws. In particular, the conjunction-disjunction distributivity is violated. There is again nothing mystical. The Boolean logic 
is just one special model of thought \cite{Boole}. It was proposed by mathematician and it spread widely, because of its clearness and matching intuition.
However, in cognitive science and psychology it was found that humans often make decisions by violating laws of Boolean logic \cite{UB} (and in particular
Bell's type inequalities \cite{LG}).
Hence, there is nothing surprising that some natural phenomena can be described by novel (non-Boolean) models of logic. Such formal logic 
was created by von Neumann \cite{VN}, see also von Neumann and Birkhoff \cite{VBR}. From this viewpoint, Bell's ``no-go theorem'' supports (indirectly 
via coupling of probability with logic) the interpretation of quantum theory as a departure from classical to special nonclassical (quantum)
logic. However, inter-relation of classical and nonclassical logics is not the main topic of this paper; we shall concentrate on probability.
For reader's convenience, we presents basics of quantum logic in the appendix.  
 
The final theme of this paper is about  a possibility to ``embed'' a noclassical probabilistic model, e.g., quantum probability, into the 
classical (Kolmogorov) probability model. From the viewpoint of the presented analogy geometry-probability, it would not be surprising if
a non-Kolmogorovian model were embedded in some (may be nonunique) way into the Kolmogorovian model. We know that non-Euclidean geometries
can be {\it modeled} by using surfaces in the Euclidean space. As was shown in \cite{t1}, \cite{t2} even in this aspect probability is similar to geometry; 
it is really possible to embed, e.g., quantum probability into the classical probability model. The embedding proposed in these papers is based 
on treatment of quantum probabilities (correlations) as conditional classical probabilities (correlations). In particular, the EPR-Bohm correlations
used in Bell's ``no-go theorem'' can be treated in this way. In this paper we present the construction proposed in \cite{t1}, \cite{t2} in the 
compact form.     

It may be useful to remark that after the appearance of the arXiv-preprint \cite{ARR} I received a few critical comments on the title of this paper, 
this is issue is discussed in section \ref{BCBC}.

\section{Methodology of classical probability theory}

The reader may be surprised that by discussing Bell's type inequalities we do not say practically any word about {\it hidden variables}, see
only section \ref{hidden}. One of the main reasons for this is that classical probability theory does not admire at all hidden parameters, may be 
opposite to classical statistical physics. For example, how does    classical probability theory describe the random experiment 
on coin tossing? Does it present the space of hidden parameters leading to various results of coin tossing? Not at all! It uses so to say 
the operational description based solely on the ``results of measurements'', head (H) or tail (T). So, the space of ``elemenraty events''
for the random experiment with $n$-times tossing consists of vectors $\omega=(x_1,...,x_n),$ where $x_j=H,T.$ We can say that classical probability 
theory does not like {\it counrefactuals.} As another example, consider the classical Brownian motion. To construct the classical probability space,
we do not try to present the space of hidden parameters generating the trajectories of Brownian particles. Such trajectories by themselves are used 
as elementary events. This is again a kind of the operational description. It involves as much as possible information which can be gained from 
measurements. Classical probabilist would never discuss results which would appear if the experimental context  were changed. Andrei Kolmogorov emphasized
that each experimental context induces its own probability space \cite{K}, see also \cite{book2}, \cite{INT}, \cite{arXiv0}. Although Kolmogorov had never discussed explicitly incompatibility 
of experimental contexts, it seems that such a possibility was completely clear for him \cite{K}.\footnote{One may be 
curious whether in classical probability the probabilistic 
results obtained for incompatible experimental contexts (in other words described by different probability spaces) are used at all. My position was that 
from the classical probability viewpoint to combine incompatible probabilistic data is meaningless; in particular,  to put correlations collected in incompatible experiments 
in a single inequality. However, at least for the first part of the previous statement my position was weaken after an exciting recent debate with Richard
Gill who pointed that in classical statistics data collected for incompatible experimental contexts are used to {\it compare outputs for 
different contexts}, see section \ref{inc} for details. (Of course, one has to have in mind that statistics and probability may differ essentially 
in their methodologies for treating of outputs of random experiments.)}  

We can say that in classical probability theory the problem of hidden variables does not arise, because the operational description based
on the measured quantities  is simpler and this is the main reason for its use. So, one simply is not interested whether hidden variables 
exist or not, since there is a possibility to use simple and self-consistent operational description. This position is surprisingly close
to Bohr's position \cite{BR_AC}. It seems that Bohr would not demonstrate any interest to Bell's inequality, since Bohr was completely fine 
with the operational representation of quantum probabilities.

Nevertheless, if one likes it is possible to treat the elementary events of  the classical probability model for violation of 
CHSH-inequality as hidden variables. However, such ``hidden variables'' cannot be represented solely by using the degrees of freedom of 
quantum systems, e.g., pairs of entangled photons. They include additional parameters determined by the experimental context, cf. \cite{}, 
namely, parameters of random generators for selections of experimental settings; in this case the pairs of angles for polarization beam 
splitters. Such hidden variables are in some sense nonlocal, since one random generator is located in one lab and another in another lab.
However, this is simply completely classical nonlocality of the experimental context which has nothing to do with 
so-called ``quantum nonlocality.'' At the same time quantum observables are represented by local classical random variables, each of them 
depends only on functioning of the random selection generator located in the lab, where this variable is measured.

We also remark that the presented methodology of treatment of random experiments is heavily explored in one special interpretation of quantum 
mechanics, namely, {\it the consistent histories interpretation,}
 see especially the monograph of Bob Griffiths \cite{Griffiths}. 
Our explanation of violation of Bell's inequality is similar to the one used in the consistent histories approach 
to quantum mechanics \cite{Griffiths}: the existence of inconsistent families of histories implies non-Kolmogorovness. However, the adherences of the  
consistent histories approach are fine with the recognition of non-Kolmogorovness, they did not constructed embeddings of inconsistent 
histories in one ``large Kolmogorov probability space.''

\section{Boolean logic and Kolmogorovian probability}
\setcounter{equation}{0}

In 19th century George Boole wrote the book \cite{Boole},  in which he formalized 
the semantics of classical logic, he also formulated the laws of probability based on this logic.

One of the most important futures of Boolean logic is that it serves as the basis of the modern probability theory \cite{K}:
the representation of events by sets, subsets of some set $\Omega,$ so called sample space, or space of elementary events,
 The system of sets 
representing events, say ${\cal F},$ matches with the operations of Boolean logics; 
 ${\cal F}$ is so-called $\sigma$-algebra of sets\footnote{Here the symbol 
$\sigma$ encodes ``countabile''. In American terminology such systems of subsets are called $\sigma$-fields.}. 
It is closed with respect to the (Boolean) operations of 
(countable) union, intersection, and complement (or in logical terms ``and'', ``or'', ``no'').  
Thus the first lesson for a physics student is that by applying any theorem of probability theory, e.g., 
the law of large numbers,  one has to be aware that Boolean logic is in the use. The set-theoretic  
model of probability was presented by Andrei Nikolaevich Kolmogorov in 1933 \cite{K}; it is based on the following two natural (from the 
Boolean viewpoint) axioms:
\begin{itemize}
\item  AK1) events are represented as elements of a $\sigma$-algebra and operations for events are described by Boolean logic;

\item AK2) probability is represented as a  probabilistic measure.

\end{itemize}

We remind that a  probabilistic measure $p$ is a (countably) additive function  on a $\sigma$-algebra ${\cal F}$: 
$$
p(\cup_{j=}^\infty A_j)= 
\sum _{j=}^\infty p(A_j) \; \mbox{for} \; A_j \in   {\cal F}, A_i \cap A_j =\emptyset , i \not= j,
$$ 
which is   valued in $[0,1]$ and normalized by 1, $p(\Omega)=1.$ 

The triple ${\cal P}=(\Omega, {\cal F}, p)$ is called (Kolmogorov) {\it probability space.}

We also remind the definition of a {\it random variable} as a measurable function,
 $a: \Omega \to \mathbf R.$  In classical probability theory random variables represent 
observables.

 {\small Here measurability has the following meaning. The set of real numbers $\mathbf R$ is endowed 
 with the {\it Borel $\sigma$-algebra} ${\cal B}:$ the minimal  $\sigma$-algebra containing all open and closed intervals. 
Then for any $A\in {\cal B}$
 its inverse image $a^{-1}(A) \in {\cal F}.$ This gives the possibility to define on ${\cal B}$ 
the probability distribution of a random variable, $p_a(A) = p(a^{-1}(A)).$}

The mathematical expectation (average) of a random variable $\xi$ is defined as its integral:
$E\xi= \int_\Omega \xi(\omega) dp(\omega).$

Finally, we point to an exceptional role which is played by 
conditional probability in the Kolmogorov model. This sort of probabilities is not derived 
in any way from ``usual probability''; conditional probability 
is {\it per definition} given by the {\it Bayes formula:}  
\begin{equation}
\label{Bayes} P(B \vert C) =  P(B\cap C)/ P(C),
 P(C)>0.
\end{equation}
By Kolmogorov's interpretation it is the \textit{probability of an event $B$
to occur under the condition that an event $C$ has occurred.} One can immediately see that this formula is one of 
strongest exhibitions of the Boolean
structure of the model; one cannot  even assign conditional probability to an event without 
using the Boolean operation of intersection.

Thus the second lesson for a physics student is that {\it probability is an axiomatic theory}, as, e.g., geometry.

\section{CHSH-inequality as a theorem of Kolmogorov probability theory}
\setcounter{equation}{0}

Let ${\cal P}=(\Omega, {\cal F}, p)$ be a Kolmogorov probability space. 
For two random variables $A$ and $B,$ we set 
$$
<A, B>=  E (AB)= \int_\Omega A(\omega) B(\omega) dp(\omega). 
$$
\medskip

{\bf Theorem 1.} (CHSH-inequality) {\it Let $A^{(i)},  B^{(j)}, i,j =1,2,$ be 
random variables with values in [-1,1]. Then the corresponding combination of correlation 
\begin{equation}
\label{COR_k7}
S= <A^{(1)},B^{(1)}> + <A^{(1)},B^{(2)}> + <A^{(2)}, B^{(1)}> - <A^{(2)},B^{(2)}>,
\end{equation}
satisfies the  CHSH-inequality:}
\begin{equation}
\label{LI77}
\vert S \vert \leq 2.
\end{equation}

{\bf Proof.} It is easy to show that for any quadruple of random variables valued in [-1,1] the following inequality holds:
$$
2 \leq A^{(1)}(\omega)B^{(1)}(\omega) + A^{(1)}(\omega)B^{(2)}(\omega) + 
A^{(2)}(\omega) B^{(1)}(\omega) - A^{(2)}(\omega)B^{(2)}(\omega) \leq 2.
$$
By integrating this inequality with respect to the probability measure $p$ we obtain (\ref{LI77}).

\section{From Euclid to Lobachevesky and from Kolmogorov to Bell}
\setcounter{equation}{0}

To understand better the axiomatic nature of the modern set-theoretic model of probability, it is useful to make comparison with another 
axiomatic theory - geometry. We can learn a lot from history of development of geometry. Of course, the biggest name in geometry is Euclid. His 
axiomatics of geometry was considered as the only possible during about two thousand years. It became so common that people started to identify 
Euclidean model of geometry with physical space. In particular, Immanuel Kant presented deep philosophic arguments \cite{Kant} that physical 
space is Euclidean. The Euclidean dogma was rejected as the result of internal mathematical activity, the study of a possibility of derivation 
of one of axioms from others. This axiom was the famous fifth postulate: through a point not on a line, there is precisely one line parallel to 
the given one. Nikolay Ivanovich Lobachevsky was the first who demonstrated that this postulate can be replaced with one of its negations.
This led him to a new geometric axiomatics, the model which nowadays is known as Lobachevsky geometry 
(or hyperbolic geometry). Thus the Euclidean geometry started to be treated as just one of possible models of geometry. And 
this discovery revolutionized first mathematics (with contributions of Gauss, Bolyai, and especially Riemann)
and then physics (Minkowski, Einstein, Hilbert).   

This geometry lesson tells us that there is no reason to expect that the Kolmogorovian model is the only possible axiomatic model 
of probability. One can expect that by playing with  the Kolmogorovian axioms as Lobachevsky played with the Euclidean ones mathematicians could create 
{\it non-Kolmogorovian models of probability} which may be useful for various applications, in particular in physics. However, in the case of 
probability the historical pathway of development of geometry was not repeated. Mathematicians did not have two thousand years to rethink 
the Kolmogorovian axiomatics... 

\section{Non-Kolmogorovian nature of quantum probability; no-go theorems}
\setcounter{equation}{0}

New physics, QM, intervened brutally in the mathematical kingdom. The probabilistic structure of 
QM did not match with classical probability theory based on the set-theoretic approach of Kolmogorov. At the first stage
of development  of QM this mismatching was not so visible. The first clink  came in the form of {\it Born's rule}: 
\begin{equation}
\label{BRULE}
p(x)= \vert \psi(x) \vert^2,
\end{equation}
where $\psi(x)$ is the wave function and $p(x)$ is the probability to detect a particle at the point $x.$ Here not the probability, 
but the wave function
is primary. What is encoded in this presence of complex amplitudes behind probabilities obtained in quantum measurements? 

The first who paid attention to peculiarity of the probabilistic structure of QM comparing with the probabilistic structure of classical 
statistical mechanics was John von Neumann \cite{VN}. In particular, he generalized Born's rule to quantum observables represented by 
Hermitian operators; 
for observable represented by an operator with purely discrete spectrum, the probability to obtain the value $\lambda$ as 
the result of measurement is given as
\begin{equation}
\label{BRULE1}
p(\lambda)= \Vert P_\lambda \psi \Vert^2,
\end{equation}
where $P_\lambda$  is the projector corresponding to the eigenvalue $\lambda.$ (Here $A= \sum_\lambda \lambda P_\lambda).$

In his seminal book \cite{VN} von Neumann pointed out that, opposite to classical statistical mechanics where 
randomness of results of measurements is a consequence of variability of physical parameters such as, e.g., the position and momentum of a classical 
particle, in QM the assumption about the existence of such parameters (for a moment may be still hidden and unapproachable by existing measurement 
devices) leads to contradiction. This statement presented in \cite{VN} is known as {\it von Neumann no-go theorem}, theorem about impossibility 
to go beyond the description of quantum phenomena based on quantum states: it is impossible to construct a theoretical model providing a finer 
description of those phenomena than given by QM.\footnote{This theorem was criticized for unphysical assumptions used by von Neumann to approach 
his no-go conclusion; especially strong critique was from the side of John Bell \cite{B}, the author of another famous no-go theorem; calmer 
critical arguments were presented by Leslie Ballentine \cite{BL}.  (We also remark that, although in the modern literature the von 
Neumann statement is called 
``theorem'', in the German eddition he called it ``ansatz''.)} Thus von Neumann was sure that it is impossible to construct a 
classical probability measure on the space of some hidden variables which would reproduce probabilities obtained in quantum measurements. 

Later this statement was confirmed by other no-go theorems, e.g., of Bell \cite{B}. 

\medskip

{\it Bell's ``no-go theorem'' says that Bell type inequalities, e.g., the CHSH-inequality  (\ref{LI77}),
which are derived in Kolmogorovian model of probability are violated for correlations calculated 
in the quantum probability model.} 

\medskip

Thus we conclude
that the former model of probability has to be rejected as inapplicable to these correlations and that the latter model (which is non-Kolmogorovian)
has to be in the use. From the viewpoint of the scientific methodology based on the transition from one mathematical model to another, this is 
a natural step of the scientific evolution. If one try to go beyond this methodology and to interpret the violation of Bell's inequality not simply 
as generating the aforementioned transition from one model of probability to another, but to attract additional elements (which do not have 
formal mathematical meaning) such as (non)locality or (un)realism, then he starts a random walk in jungles of arguments and counter-arguments.
And this happened in debates on Bell's argument.   

\section{Kolmogorovization of quantum probability}
\setcounter{equation}{0}

\subsection{Embedding: Geometry-probability analogy}

As we pointed out in the introduction, the geometry-probability analogy stimulates us to construct embeddings of the quantum model 
of  probability into the classical probability model. On the other hand, two models of probability seems to be so different that 
such embeddings are impossible. In the classical model events are represented by sets and operations for events are based on  
classical logic, in the quantum model events are represented by subspaces and operations for them violate the laws of 
classical logic, they are based on quantum logic of von Neumann-Birkhoff, see the appendix. In the classical model probability 
is given by a measure on a $\sigma$-algebra of events and in the quantum model it is given by Born's rule, by the operator trace.
However, again the geometry-probability analogy tells us that such differences may be not a problem. 

We know that a non-Euclidean geometry having very exotic features (from the Euclidean viewpoint) can be modeled by using surfaces 
in the Euclidean space. Let us turn again to the Lobachevskian geometry; consider, for example, the so called hyperbolic plane.   
There are various models of this plane based on the Euclidean plane. Consider, for example, 
the Poincar\'e disc model. It is based the interior of a circle (in the Euclidean plane) and lines are represented by arcs of 
circles that are orthogonal to the boundary circle, plus diameters of the boundary circle. We stress that, although in this representation 
 the hyperbolic plane is given by a domain of the Euclidean plane, the correspondence between the basic entities of the models is 
nontrivial. Lines of the hyperbolic geometry are not at all straight lines of the Euclidean geometry. Therefore if we hope to embed 
the quantum model of probability  into the classical model of probability, we have to be prepared that the quantum probability will be 
represented in some nontrivial way.   

As was pointed out in the introduction, besides geometry-probability similarities, there are also differences. Before to proceed to 
embedding of quantum probability into classical probability, we discuss one specialty of the Kolmogorov model. 

\subsection{Kolmogorov model: Coupling with experiment}

The quantum model describes {\it all possible quantum measurements}. We can always select a Hilbert state space such that 
all possible quantum observables are represented by Hermitian operators acting in this space (more generally by POVMs).  
This representation of quantum observables, all in one Hilbert space, is definitely a mathematical idealization. 
Here even incompatible observables peacefully coexist, although the joint probability distributions are well defined 
only for groups of {\it compatible observables.}   

Surprisingly the classical model is closer to real experimental situation\footnote{Although typically the opposite is claimed, namely, that 
the quantum model is an operational model designed just to describe results of measurements, cf. also with the operational derivations 
of QM in the spirit of D' Ariano et al. \cite{dariano}}. Here we do not try to construct some huge 
Kolmogorov space, i.e., a triple ${\cal P}=(\Omega, {\cal F}, p),$  such that all possible observables would
be represented by random variables on this space, measurable functions $a: \Omega \to \mathbf R.$ As was emphasized 
by Kolmogorov in section 2 of his monograph \cite{K}, each experimental context $C$ is represented by its own probability 
space ${\cal P}_{C}=(\Omega_{C}, {\cal F}_{C}, p_{C}),$. (This message was practically washed out from modern textbooks in probability.) 
Of course, sometimes one needs 
to operate with data corresponding to a few experimental contexts $C_1,....,C_N.$ They are represented by the probability 
spaces    ${\cal P}_{C_j}=(\Omega_{C_j}, {\cal F}_{C_j}, p_{C_j}).$ How can one construct the ``unifying probability space''?
The procedure is known as randomization. First of all one have to decide how often each experimental context will be in the use and to
assign probabilities $q_1,..., q_N$ to contexts-realizations. This is an important step. One has to take into account 
the randomness of selection of experimental contexts. This randomness is real and it cannot be ignored. Then  by selecting
the probability spaces ${\cal P}_{C_j}$ with the probabilities $q_j$
a ``big probability space'' is constructed.

This fundamental coupling of the classical probability model to experiment has to be taken into account if one wants to represent 
 quantum probability in a classical probability space. In the light of the above discussion it is natural to construct 
 the classical probabilistic representation for any finite group of quantum measurements. We shall consider the most interesting example,
the test to violate the CHSH-inequality (\ref{LI77}). (Since here we have experimental verification of violation under the assumption of fair sampling.)  

\section{Classical probability model for quantum correlations violating the CHSH inequality}

Now we present embedding of the probabilities (and correlations) for joint measurements of polarizations
for pairs of photons given by QM and violating 
the CHSH-inequality (\ref{LI77}) into a Kolmogorov probability space.
This construction was used in \cite{t1}, \cite{t2}. Here we present it in a clearer way.   

To verify an inequality of this type, one should put in it statistical data collected
for {\it four pairs of settings } of polarization beam splitters (PBSs):
$$
\theta_{11}=(\theta_1, \theta_1^\prime), \theta_{12}=(\theta_1,
\theta_2^\prime), \theta_{21}=(\theta_2, \theta_1^\prime),
\theta_{22}=(\theta_2, \theta_2^\prime).
$$
Here $\theta= \theta_1, \theta_2$ and $\theta^\prime=
\theta_1^\prime, \theta_2^\prime$ are selections of angles  for orientations of respective PBSs.
The selection of the angle $\theta_i$ determines the corresponding polarization observable,
$
 a_{\theta_i}= \pm 1.
$
There are two detectors coupled to the PBS with the fixed
$\theta$-orientation: ``up-polarization'' detector
and  ``down-polarization'' detector.  
A click of the up-detector assigns to the random variable $a_{\theta}(\omega)$
the value +1  and a click of the down-detector assigns to it the
value -1. In the same way selection of the angle $\theta^\prime$ determines the corresponding polarization observable,
$
b_{\theta_i^\prime}= \pm 1.
$
Thus, in fact, the CHSH-test consists of four different experiments corresponding to settings $\theta_{ij}.$ 
Our aim is to unify these four experiments into a single experiment with random selection of experimental 
settings. In principle, such unification is used in modern tests of the CHSH-inequality in which settings are 
selected with the aid of random generators. 

For the illustrative purpose, it is more useful to map this experiment 
with random switching of orientations of two fixed PBSs onto the experiment in which all settings are unified at the 
``hardware level'', i.e., the experiment with 4 PBSs oriented with the angles $\theta_1, \theta_2$ and $\theta^\prime_1, \theta^\prime_2$
and each PBS is equipped with its own two detectors, so there are totally 8 detectors.  

Such an experimental scheme was used in the pioneer experiment of A. Aspect \cite{Aspect} 
with one difference: he used single channel PBSs.  We, finally present the corresponding citation  
of  Aspect \cite{Aspect1}, see also \cite{Aspect},  section ``Difficulties of an ideal experiment'':

``We have done a step towards such an ideal experiment by using the modified scheme shown on Figure 15.
In that scheme, each (single-channel) polarizer is replaced by a setup involving a
switching device followed by two polarizers in two different orientations: $a$ and $a^\prime$ on side
I, $b$ and $b^\prime$ on side II. The optical switch $C1$ is able to rapidly redirect the incident light
either to the polarizer in orientation $a,$ or to the polarizer in orientation $a^\prime.$ This setup is
thus equivalent to a variable polarizer switched between the two orientations $a$ and $a^\prime.$ A
similar set up is implemented on the other side, and is equivalent to a polarizer switched
between the two orientations $b$ and $b^\prime.$ In our experiment, the distance $L$ between the two
switches was 13 m, and $L / c$ has a value of 43 ns. The switching of the light was effected by home built devices, based on the
acousto-optical interaction of the light with an ultrasonic standing wave in water. The
incidence angle (Bragg angle) and the acoustic power, were adjusted for a complete
switching between the 0th and 1st order of diffraction.''

Figure 15 can be found in  \cite{Aspect1}, p.26. The only difference of our scheme that each of four PBSs has two output channels. 

\begin{figure}[htpb]
\centering{}
\includegraphics[scale=1]{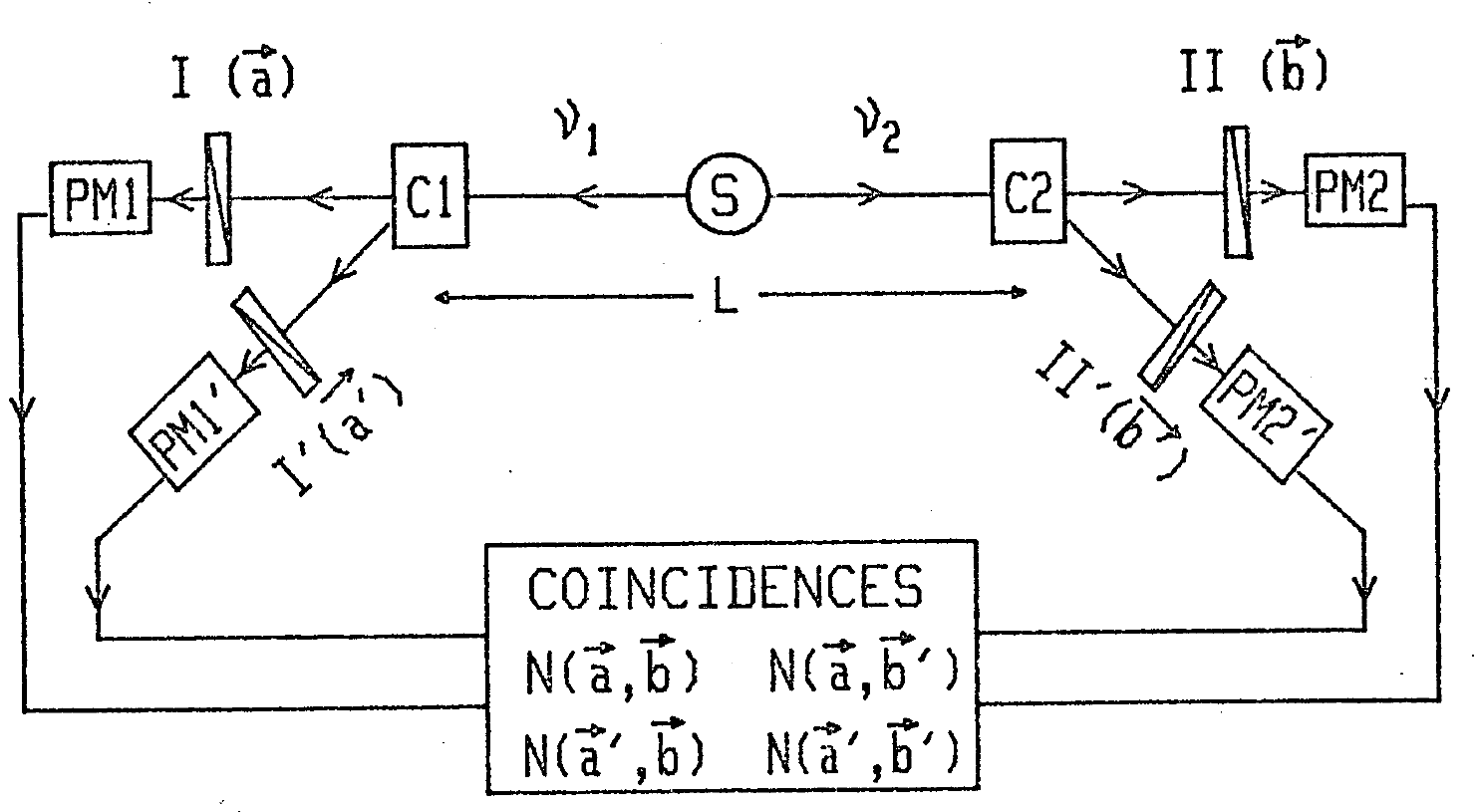} 
\caption{The scheme of the pioneer experiment of A. Aspect with four  beam splitters \cite{Aspect}.}
\label{fig1_asp} 
\end{figure}

\subsection{Experiment taking into account random choice of settings}
\label{4_8}

\begin{itemize}
\item[a).] There is a source of entangled photons. 

\item[b).] There are 4 PBSs and corresponding pairs of detectors for each PBS, totally 8 detectors. PBSs are
labeled   as $i= 1,2$ (at the left-hand side, LHS) and $j=1,2$ (at the right-hand side, RHS). 

\item[c).] Directly after source there are 2 distribution devices, one at LHS and one at RHS.
At each instance of time, $t=0, \tau, 2\tau, \ldots$ each device opens 
the port to only one (of two) optical fibers going to  the
corresponding two PBSs. For simplicity, we suppose that each pair
of ports $(i,j),$ 
$(1,1), (1,2), (2,1), (2,2),$ can be opened with equal
probabilities\footnote{In general  for selection of experimental settings  we can use any probability distribution, 
$ \sum_{i=1, j=1}^2 {\bf P}(i,j)=1.$ Under the natural assumption that the 
distribution devices operate independently, we introduce two probability distributions 
$P_L(i), i=1,2, P_L(1)+P_L(2)=1,$ and   
$P_R(j), j=1,2, P_R(1)+P_R(2)=1,$ and set ${\bf P}(i,j)= P_L(i) P_R(j).$
}:
$$
{\bf P}(i,j)=1/4.
$$
\end{itemize}

 Now we introduce the observables measured in this experiment. They are modifications of 
 the polarization observables $a_{\theta_i}, i=1,2,$ and $b_{\theta^\prime_j}, j=1,2.$ We start 
 with the ``LHS-observables'':

1) $A^{(i)} = \pm 1, i=1,2$ if the corresponding (up or
down)  detector is coupled to $i$th PBS (at LHS) fires and the $i$-th channel is open;

2) $A^{(i)}= 0$ if the $i$-th channel (at LHS) is  blocked. 

In the same way we define the ``RHS-observables'' $B^{(j)}=0, \pm 1,$
corresponding to PBSs $j=1,2$.

Thus unification of 4 incompatible experiments of the CHSH-test into a single experiment modifies the range of values
of polarization observables for each of 4 experiments; the new value, zero, is added to reflect the random choice of experimental 
settings. We emphasize that this value has no relation to the efficiency of detectors. In this model we assume that detectors have 
100\% efficiency. The observables take the value zero when the optical fibers going to the corresponding PBSs are blocked.

\subsection{Kolmogorov space for incompatible observables}

Our aim is to construct a proper Kolmogorov probability space for the
experiment which was described in the previous section. 
In fact, we shall present a general construction for
combining probabilities produced by a few 
experiments which can be incompatible.

In the CHSH-test we operate with probabilities $p_{ij}(\epsilon,
\epsilon^\prime), \epsilon, \epsilon^\prime= \pm 1,$ -- to get the results
$a_{\theta_i}=\epsilon, b_{\theta_{j}^\prime}=\epsilon^\prime$ in
the experiment with the fixed pair of orientations $(\theta_i,
\theta_j^\prime).$ From QM we know that, for the singlet state,   
these EPR-Bohm probabilities are given by expressions: 
\begin{equation}
\label{QM} 
p_{ij}(\epsilon, \epsilon)=\frac{1}{2} \cos^2
\frac{\theta_i-\theta_j^\prime}{2}, p_{ij} (\epsilon,
-\epsilon)=\frac{1}{2} \sin^2 \frac{\theta_i-\theta_j^\prime}{2}.
\end{equation}
However, this special form of probabilities is not important for
us. Our construction of unifying Kolmogorov probability space
works well for any collection of probabilities 
\begin{equation}
\label{QM1} 
p_{ij}>0: \sum_{\epsilon, \epsilon^\prime} p_{ij} (\epsilon,
\epsilon^\prime)=1. 
\end{equation}
Thus we proceed in this general situation, (\ref{QM1}). 
Hence, we consider the experiment described in section \ref{4_8} and producing some collection of probabilities $(p_{ij}).$
The special choice of the  probabilities (\ref{QM}) will be used only to concrete considerations and to violate
the CHSH-inequality, see section \ref{TT7}.  
 
Let us now consider the set of points $\Omega$ (the space of ``elementary events'' in Kolmogorov's terminology): 
$$
\Omega= \{(\epsilon_1, 0, \epsilon_1^\prime, 0), (\epsilon_1, 0, 0,
\epsilon_2^\prime), (0, \epsilon_2, \epsilon_1^\prime, 0), (0,
\epsilon_2, 0, \epsilon_2^\prime)\},
$$
where $\epsilon =\pm 1, \epsilon^\prime= \pm 1.$
These points correspond to the following events:
e.g., $(\epsilon_1, 0, \epsilon_1^\prime, 0)$
means: at LHS the PBS with $i=1$ is coupled and  the PBS with $i=2$ is uncoupled and the same situation  is 
at RHS: the PBS with $j=1$ is coupled and  the PBS with $j=2$ is uncoupled; the result of measurement at LHS
after passing the PBS with $i=1$ is given by $\epsilon_1$ and at RHS  by $\epsilon_1^\prime.$

We define the following probability measure\footnote{To match completely with Kolmogorov's terminology, we have to select a $\sigma$-algebra ${\cal F}$ of subsets of $\Omega$ representing events and define 
the probability measure on ${\cal F}.$ However, in the case of a finite $\Omega=\{\omega_1,..., \omega_k\}$ the system of events ${\cal F}$ is always chosen as consisting 
of all subsets of $\Omega.$ To define a probability measure on such ${\cal F},$ it is sufficient to define it for one-point sets, 
$(\omega_m) \to {\bf P}(\omega_m), \sum_m {\bf P}(\omega_m)=1,$ and to extend it by additivity:
 for any subset $O$ of $\Omega, {\bf P}(O)= \sum_{\omega_m\in O} 
{\bf P}(\omega_m).$  
}
on $\Omega$
$${\bf P}(\epsilon_1, 0, \epsilon_1^\prime, 0) = \frac{1}{4} p_{11}(\epsilon_1, \epsilon_1^\prime),
{\bf P}(\epsilon_1, 0, 0, \epsilon_2^\prime) = \frac{1}{4}
p_{12}(\epsilon_1, \epsilon_2^\prime)
$$
$$
{\bf P}(0, \epsilon_2, \epsilon_1^\prime, 0) = \frac{1}{4}
p_{21}(\epsilon_2, \epsilon_1^\prime), {\bf P}(0, \epsilon_2, 0,
\epsilon_2^\prime) = \frac{1}{4} p_{22}(\epsilon_2,
\epsilon_2^\prime).
$$

We now define random variables $A^{(i)}(\omega), B^{(j)}
(\omega):$
\begin{equation}
\label{QMA} 
A^{(1)}(\epsilon_1, 0, \epsilon_1^\prime, 0)= A^{(1)}(\epsilon_1,
0, 0, \epsilon_2^\prime)= \epsilon_1,
A^{(2)}(0, \epsilon_2, \epsilon_1^\prime, 0)= A^{(2)}(0, \epsilon_2, 0, \epsilon_2^\prime)= \epsilon_2;
\end{equation}
\begin{equation}
\label{QMB} 
B^{(1)}(\epsilon_1, 0, \epsilon_1^\prime, 0)= B^{(1)}(0,
\epsilon_2, \epsilon_1^\prime, 0)= \epsilon_1^\prime,
B^{(2)}(\epsilon_1, 0, 0, \epsilon_2^\prime)= B^{(2)}(0,
\epsilon_2, 0, \epsilon_2^\prime)=\epsilon_2^\prime.
\end{equation}
and we put these variables equal to zero in other points. 

\medskip

{\bf Remark.} (Locality of the model) 
We remark that the values of the  LHS-variables $A^{(1)}(\omega), A^{(2)}(\omega)$ 
depend only the first two coordinates of $\omega$ and   the values of the  
RHS-variables $B^{(1)}(\omega), B^{(2)}(\omega)$ 
depend only on the last two coordinates of $\omega.$ 
Thus the value of $A^{(i)}(\omega)$ does not depend 
on the values of  $B^{(j)}(\omega).$ They neither depend on the 
action of the RHS distribution device; for the LHS-variable it is 
not important which port is open or closed by the RHS distribution 
device. The RHS-variables  $B^{(j)}(\omega)$ behave in the same way.
Thus the random variables under consideration are determined {\it locally.}
  
\medskip  
  
We find two dimensional probabilities $${\bf P}(\omega \in \Omega:
A^{(1)} (\omega)= \epsilon_1, B^{(1)}(\omega)= \epsilon^\prime_1)
={\bf P}(\epsilon_1, 0, \epsilon_1^\prime, 0)= \frac{1}{4} p_{11}
(\epsilon_1, \epsilon_1^\prime), \ldots,
$$
$$
{\bf P} (\omega \in \Omega: A^{(2)} (\omega)= \epsilon_2,
B^{(2)}(\omega)= \epsilon_2^\prime) = {\bf P}(0, \epsilon_2, 0,
\epsilon_2^\prime) =\frac{1}{4} p_{22}
(\epsilon_2, \epsilon_2^\prime).
$$

We also consider the random variables which are responsible for
selections of pairs of ports to PBSs. For the device at LHS: 
$$
\eta_L(\epsilon_1, 0, 0,\epsilon_2^\prime)= \eta_L(\epsilon_1, 0, \epsilon_1^\prime, 0)= 1,
\eta_L(0, \epsilon_2, 0, \epsilon_2^\prime)= \eta_L(0, \epsilon_2, \epsilon_1^\prime, 0)= 2 .
$$
For the device at RHS: 
$$
 \eta_R(\epsilon_1, 0, \epsilon_1^\prime, 0)= \eta_R(0, \epsilon_2,\epsilon_1^\prime, 0)= 1, 
\eta_R(0, \epsilon_2, 0, \epsilon_2^\prime)=  \eta_R(\epsilon_1, 0, 0,\epsilon_2^\prime)= 2.
$$

\subsection{Validity of CHSH-inequality for correlations taking into account randomness 
of selection of experimental settings}  

Consider the correlations with respect to the probability ${\bf P}$ (which takes into account 
randomness of selections of experimental settings), 
\begin{equation}
\label{COR}
<A^{(i)}, B^{(j)}>= \int_\Omega A^{(i)}(\omega) B^{(j)}(\omega) d{\bf P}(\omega).
\end{equation}
These are classical correlations for random variables taking values in [-1,1] and for them the CHSH-inequality $|S| \leq 2,$
see (\ref{LI77}),  holds (Theorem 1).  Here $S$ is defined by (\ref{COR_k7}).

We remark that 
\[
<A^{(i)}, B^{(j)}> = \frac{1}{4}\Big[\sum_{\epsilon_1, \epsilon_1^\prime= \pm 1} 
A^{(i)}(\epsilon_1, 0, \epsilon_1^\prime, 0)) B^{(j)}(\epsilon_1, 0, \epsilon_1^\prime, 0)) p_{11}(\epsilon_1, \epsilon_1^\prime)
\]
\[
+\sum_{\epsilon_1, \epsilon_2^\prime= \pm 1} A^{(i)}(\epsilon_1, 0, 0,\epsilon_2^\prime) B^{(j)}
(\epsilon_1, 0, 0,\epsilon_2^\prime)  p_{12}(\epsilon_1, \epsilon_2^\prime)
\]
\[
+  \sum_{\epsilon_2, \epsilon_1^\prime= \pm 1} 
A^{(i)} (0, \epsilon_2, \epsilon_1^\prime, 0) B^{(j)}(0, \epsilon_2, \epsilon_1^\prime, 0)  p_{21}(\epsilon_2, \epsilon_1^\prime)
\]
\[
+ \sum_{\epsilon_2, \epsilon_2^\prime= \pm 1} A^{(i)}(0,\epsilon_2, 0, \epsilon_2^\prime)B^{(j)}(0,\epsilon_2, 0, \epsilon_2^\prime) p_{22}(\epsilon_2, \epsilon_2^\prime) \Big].
\]
Let us  fix some setting, e.g., $i=1, j=1.$ Then taking into account the definitions of $A^{(1)}$ and $B^{(1)}$ we found that
in the above expression only the first summand is nonzero.Thus
$<A^{(1)}, B^{(1)}> =$
\[
 \frac{1}{4} \sum_{\epsilon_1, \epsilon_1^\prime= \pm 1} 
A^{(i)}(\epsilon_1, 0, \epsilon_1^\prime, 0)) B^{(j)}(\epsilon_1, 0, \epsilon_1^\prime, 0)) p_{11}(\epsilon_1, \epsilon_1^\prime)=
\frac{1}{4}\sum_{\epsilon_1, \epsilon_1^\prime= \pm 1} \epsilon_1 \epsilon_1^\prime p_{11}(\epsilon_1, \epsilon_1^\prime).
\]
Thus classical correlations (taking into account randomness of setting selections) 
coincide (up to the factor $1/4)$ with the correlations corresponding to the probability measures $(p_{ij}):$ 
\begin{equation}
\label{LI}
 C_{ij} \equiv \sum_{\epsilon_i, \epsilon_j^\prime= \pm 1} \epsilon_i \epsilon_j^\prime p_{ij}(\epsilon_i, \epsilon_j^\prime),
\end{equation}
namely,
\begin{equation}
\label{LI1}
C_{ij}= 4 <A^{(i)}, B^{(j)}>.
\end{equation}
In particular, we can select the probabilities $(p_{ij})$ as the EPR-Bohm probabilities for the singlet state, see (\ref{QM}),
then $C_{ij}$ are the EPR-Bohm correlations used to violate the CHSH-inequality. 

We stress that the {\it correlations $C_{ij}$ are larger than classical ones}.
In the general case, see footnote ?, 
\begin{equation}
\label{LI2}
C_{ij}= \frac{1}{{\bf P}(i,j)} <A^{(i)}, B^{(j)}>.
\end{equation}
Thus randomization washes out a part of correlation. However, this washing effect is the probabilistic reality: in any CHSH-test 
experimentalists have to determine selection of experimental settings.\footnote{ If one ignores this fact, then  his/her description of 
the CHSH-test would not match the real experimental situation; an important part of randomness involved in the experiment would be missed.
The specialty of this sort of randomness is that it is not present in the formalism of QM. The experimenters decide which random variables 
$\eta_L$ and $\eta_R$ are selected to choose the experimental settings. This situation may be disturbing for those who try to describe 
the whole experiment by QM. One has either to recognize that a part of randomness involved in the CHSH-test is not represented in the quantum
formalism or to ignore this randomness by considering simply the correlations (\ref{LI}) computed for the fixed experimental settings 
without taking into account 
randomness of selection of settings.}

\subsection{Quantum correlations as conditional classical correlations} 
 
In classical probability theory one uses the notion of {\it conditional
expectation} of a random variable (under the condition that some
event occurred). This notion is based on Bayes' formula defining conditional probability,
see (\ref{Bayes}). We shall use this notion to definite conditional correlation --
under the condition that the fixed pair $(i,j)$ of experimental settings is selected. 

Let $(\Omega, {\cal F}, {\bf P})$ be an arbitrary probability
space and
 let $\Omega_0 \subset \Omega,  \; \Omega_0 \in {\cal F,} \; {\bf P}(\Omega_0) \ne 0.$ We also consider an arbitrary random
variable $\xi:\Omega \to {\bf R}.$ Then the conditional mathematical expectation (average) of the random variable $\xi,$
conditioned to the event $\Omega_0,$ is defined as follows: 
$$
E(\xi|\Omega_0)= \int_\Omega \xi(\omega) d{\bf
P}_{\Omega_0}(\omega),
$$
where the conditional probability ${\bf P}_{\Omega_0}$ is defined by Bayes'
formula (\ref{Bayes}):
$
{\bf P}_{\Omega_0} (U) \equiv {\bf P}(U|\Omega_0)= {\bf P}(U \cap
\Omega_0)/{\bf P}(\Omega_0).
$
Let us come back to our unifying probability space. Take in the above 
definition 
$$\Omega_0\equiv \Omega_{ij}= \{ \omega \in  \Omega: \eta_L(\omega)=i, 
\eta_R(\omega)=j\}.$$ We remark that ${\bf P}(\Omega_{ij})= 1/4.$ 
The latter implies that 
\begin{equation}
\label{COR_COND}
E(A^{(i)} B^{(j)}|\eta_L=i, \eta_R=j)= \int_\Omega A^{(i)}(\omega) B^{(j)}
(\omega) d {\bf P}_{\Omega ij} (\omega) 
\end{equation}
$$
= 4\int_{\Omega_{ij}}
A^{(i)}(\omega) B^{(j)} (\omega) d {\bf P}(\omega)= 4\int_{\Omega}
A^{(i)}(\omega) B^{(j)} (\omega) d {\bf P}(\omega)
=4 <A^{(i)}, B^{(j)}>
$$
(since the product $A^{(i)}(\omega) B^{(j)} (\omega)$ is nonzero only on the set $\Omega_{ij}$).
Hence, we have:
\begin{equation}
\label{LI3}
E(A^{(i)} B^{(j)}|\eta_L=i, 
\eta_R=j) \equiv 4 <A^{(i)}, B^{(j)}>. 
\end{equation}
By comparing (\ref{LI1}) and (\ref{LI3}) we obtain that
the correlations $C_{ij}$ which correspond to the collection 
of probabilities $(p_{ij})$
coincide with the classical conditional correlations, condition with respect to the 
choice of experimental setting.

Thus,  we can identify the correlations $C_{ij}$ obtained with the aid of probabilities $(p_{ij})$ 
with the corresponding conditional correlation for the unifying Kolmogorov space:  
\begin{equation}
\label{LI4}
 C_{ij}=  E(A^{(i)} B^{(j)}|\eta_L=i, \eta_R=j).
\end{equation}
We were able to embed the correlations $C_{ij}$ collected (separately) for in general incompatible experimental settings
into the classical probability space. We remark that in principle by themselves these correlations can have the
non-Kolmogorovian structure. It can happen that there is no a single classical probability space in which 
$C_{ij}$  can be considered as unconditional correlations. 

In particular, we can select the probabilities $(p_{ij})$ as the EPR-Bohm probabilities for the singlet state, see (\ref{QM}). Then we obtain 
the {\it representation of the corresponding quantum correlations as classical conditional correlations.}

Hence, the quantum correlations are present in our classical probability model for the CHSH-test, but they are not simply correlations:
they are  conditional conditional; cf. with  the embedding of the hyperbolic plane (the Poincar\'e model) into the Euclidean place,
the lines of the hyperbolic plane are not simply Euclidean lines.  

\subsection{Violation of the CHSH-inequality for classical conditional correlations}
\label{TT7}

As we have seen, the CHSH-inequality is satisfied for classical (uncoditional) correlations (\ref{COR}). 

\medskip

{\it Is there any reason to expect that it is also satisfied for classical conditional correlations?}
 
\medskip
The answer is no. In the classical probability model Theorem 1 can be proven only for unconditional correlations. Thus in principle
conditional correlations can violate the CHSH-inequality. And this is not surprising from the Kolmogorovian viewpoint.
Set 
\begin{equation}
\label{LI5}
S_{\rm{C}}= E(A^{(1)} B^{(1)}|\eta_L=1, \eta_R=1) + E(A^{(1)} B^{(2)}|\eta_L=1, \eta_R=2)
\end{equation}
$$
 + E(A^{(2)} B^{(1)}|\eta_L=2, \eta_R=1) -
 E(A^{(2)} B^{(2)}|\eta_L=2, \eta_R=2)
$$
(here ``C'' is the abbreviation for ´´conditional'').
By using the equality (\ref{LI3}) we obtain that 
\begin{equation}
\label{LI6}
S_{\rm{C}} =4 S. 
\end{equation}
Since by general Theorem 1 the quantity $S$ is majorated by 2, the  quantity 
$S_{\rm{C}}$ is 
majorated by 8.  But the upper bound 8 is really too rough
and, to obtain a better upper bound,  we have to proceed more carefully. 

\medskip

{\bf Theorem 2.} (``Strong CHSH-inequality'') {\it Let $A^{(i)},  B^{(j)}, i,j =1,2,$ be 
random variables defined as (\ref{QMA}), (\ref{QMB}). Then the corresponding combination of correlation $S,$
see (\ref{COR_k7}), 
satisfies the stronger version of the CHSH-inequality: }
\begin{equation}
\label{LI7}
\vert S \vert \leq 1.
\end{equation}

{\bf Proof.} Consider classical correlation
$<A^{(i)}, B^{(j)}>=\int_{\Omega}A^{(i)}(\omega) B^{(j)} (\omega) d {\bf P}(\omega).$
We state again that the product $A^{(i)}(\omega) B^{(j)} (\omega)$ is nonzero only on the set $\Omega_{ij}$ and 
${\bf P} (\Omega_{ij})= 1/4.$ We also use the condition that all random variables are valued in [-1,1]. Hence, in fact, 
$$
\vert <A^{(i)}, B^{(j)}>\vert=\Big\vert
\int_{\Omega_{ij}} A^{(i)}(\omega) B^{(j)} (\omega) d {\bf P}(\omega) \Big\vert\leq {\bf P} (\Omega_{ij})= 1/4.
$$
Thus each correlation in the combination $S$ of four correlations is bounded by 1/4. Hence, the inequality (\ref{LI7})
holds.

\medskip

We remark that, for this concrete Kolmogorov space, the CHSH-inequality is ``trivialized''; in particular, the signs of correlations
in $S$ do not play any role. We remark that this theorem is valid for any collection of probabilities $(p_{ij}),$ see (\ref{QM1}).    

\medskip

{\bf Corollary 1.} (CHSH-inequality for conditional correlations.) {\it Let $A^{(i)},  B^{(j)}, i,j =1,2,$ be 
random variables defined  as (\ref{QMA}), (\ref{QMB}). Then the corresponding combination of conditional correlation $S_{\rm{C}}$ 
satisfies the inequality: }
\begin{equation}
\label{LI8}
\vert S_{C} \vert \leq 4.
\end{equation}

Take now probabilities given by (\ref{QM}), the EPR-Bohm probabilities for the singlet state. For them
$S_{C}= 2\sqrt{2}$ and the inequality (\ref{LI8}) is not violated. 

\section{Discussion}

\subsection{On the use of data from incompatible experimental contexts in statistics}
\label{inc}

Our previous analysis of the CHSH-experiment showed that one has to be very careful by operating with statistical data 
collected in incompatible ``sub-experiments'' of some ``compound experiment.'' However, we do not claim that one cannot 
use such incompatible data to derive some statistical conclusions. As was pointed by Richard Gill (email exchange) data 
collected for incompatible experimental contexts is widely used in statistics, especially in medicine:

``For instance we compare the health of non-smokers who are passively exposed to cigarette smoke (living with a smoking partner), 
with those who don't. The conclusion is that a lifetime of exposure roughly doubles your risk of lung cancer and 
other smoking related negative outcomes:  doubled from something tiny, to something also tiny.'' 

Two experimental contexts, $C_1:$ ``living with a smoking partner'', and $C_2:$ ``not'', are incompatible, we are not able 
to collect statistics by using the same person in both contexts. Nevertheless, we compare the probability distributions
$p(\cdot \vert C_i), i=1,2,$ to come to conclusions. Is such a practice acceptable from our viewpoint?  Definitely yes.
We can do everything with these contextual probabilities, but we should not forget that they are contextual, i.e., they are 
not so to say ``absolute Kolmogorovian probabilities.'' 

In the same way in the CHSH-experiment, we can compare the probabilities $p_{ij}(\epsilon, \epsilon^\prime)$ and say that , for 
one pair of angles (one experimental context) the probability of, e.g., the $(+,+)$-configuration is larger than for another pair  
(another context).

Even in medical experiments we can apply the construction of the underlying Kolmogorov space which was constructed for the 
CHSH-data. Only in this space we can work with probabilities by using the laws of classical probability theory, but absolute 
probabilities are less than contextual-conditional, this is the result of taking into account randomness of selections of the 
contexts $C_1$ and $C_2.$ This should be remembered. And it seems that statisticians remembered this difference between 
``absolute'' and conditional probabilities. (May be I wrong, but then those who treated $p(\cdot \vert C_i), i=1,2,$ as 
``absolute probabilities'' might (but need not) get the problems which are similar to the problems generated by violation of Bell's inequality 
for quantum systems.) 

We also point out that, although in medical statistics the data from incompatible experimental contexts is widely used (as Richard Gill stressed),
in another domain (where statistical methods also play the crucial role), namely, cognitive psychology and related studies in game theory and 
economics, the similar use of such data is considered as leading to ambiguous 
conclusions and paradoxes \cite{UB}, \cite{w1}-\cite{w2a}, see \cite{UB} for contextual probabilistic analysis of the situation. 
 This situation is enlighten well in games
of the {\it Prisoner's Dilemma} type. There are the following incompatible contexts: $C_+$ ($C_-)$:  
one prisoner knows that another will cooperate (not) with her and $C:$ she has no information about the plans of another 
prisoner \cite{UB}. In cognitive psychology and behavioral economics there was collected a plenty of statistical data for such games in different setups. As was pointed
out, for psychologists comparison of such data led to paradoxes. We remark that non-Kolmogorovness of data collected for the contexts 
$C_+, C_-, C$ is demonstrated via {\it violations of the formula of total probability.} This formula, as well as Bell's inequality, can be considered as a test
of non-Kolmogorovness. I proposed to use this test in cognitive science, psychology, and economics in \cite{Y}. The first experimental study 
was devoted to recognition of ambiguous figures, see Conte at al.  \cite{w3}. Later Jerome Bussemeyer coupled violation of the formula of 
total probability with data collected in games of the Prisoner's Dilemma type and with violation of 
the {\it Savage Sure Thing Principle} \cite{BusemeyerWangTownsend2006}.  The latter axiomatizes 
the rationality of players, in particular, of agents acting at the market. Generally in cognitive psychology non-Kolmogorovness was coupled to 
{\it irrationality of players} or more generally the presence of some {\it bias.}       
We also remark that in cognitive psychology Bell's type inequalities are also used as statistical tests of non-Kolmogorovness, see Conte et al. 
\cite{w4},  Asano et al. \cite{LG}.

\subsection{Consistent histories}

The consistent histories interpretation of quantum mechanics provides a similar viewpoint on violation of Bell's inequality \cite{Griffiths}.

This approach (opposite to the majority of other approaches) is heavily based on the use of the Kolmogorov axiomatics of probability theory 
in quantum physics. The adherents of consistent histories proceed at the mathematical level of rigorousness, they ``even'' define 
explicitly the spaces of elementary events serving different quantum experiments  \cite{Griffiths} (which is very unusual for other approaches).

If a set of histories is consistent, then it is possible to introduce a probability measure on it, i.e., to use the Kolmogorov probability 
model. The crucial point is that in quantum physics there exist inconsistent families of histories, so quantum measurements cannot be 
described by a single probability space; quantum probability theory is non-Kolmogorovian.  The violation of Bell's inequality is
 explained in the same way as in the present paper as a consequence of the non-Kolmogorovian structure of experimental data collected 
for different experimental contexts. In particular, the nonlocality issue is irrelative(as well as 
in the present paper). However, the consistent histories approach stops 
at this point, i.e., recognition of non-Kolmogorovness. I proceeded further and showed that, in fact, quantum probabilities can be considered as 
the classical conditional probabilities and that the Kolmogorov model covers even the quantum probabilities.

Finally, we remark that in the consistent histories approach there are no conterfactuals as well as in our approach. 

\subsection{Hidden variables}
\label{hidden}

Our model of embedding of the quantum probabilities in the Kolmogorov model 
can be considered as an extension of the space of hidden variables to include
parameters generating selections of experimental settings. Such a hidden variable
depends on the parameters for the selections of angles at both labs. One can say that
a hidden variable is nonlocal (although observed quantities are local). However, 
this nonlocal structure of a hidden variable reflects the nonlocal setup of the experiment,
and nothing else.          

\subsection{On the contribution of Bell to foundations of probability}
\label{BCBC}

It is clear that not everybody would agree to call Bell as the {\it Copernicus of Probability.} In particular, after the publication of 
the preprint \cite{ARR}, I got a few comments such as, e.g., 

\medskip

{\it ``Regarding the title of your work, ``Bell as the Copernicus of Probability'', I
think you may be overdoing things.  Whereas I very much admire Bell's work and
think he made some very important contributions to quantum foundations, he did
not solve its fundamental measurement problem (as is evident from his "Against
Measurement"), and so far as I can tell he had not the slightest idea of how to
apply standard (Kolmogorov) probability, or some modification of it, to resolve
the quantum mysteries.  He did not provide the inspiration for my CH ideas, and
I think that is also true for my colleagues who were in on its development.  He
did apply ideas of probability to draw conclusions about (classical) hidden
variables, but that seems to rather different from Copernicus' putting the sun
at the center of the solar system, an idea that did work (even if Copernicus'
formulation of it left something to be desired.)  I am not sure you want to
call Bell the Ptolemy of Probability, though that might be more accurate.
While I think it appropriate to honor great men for their achievements, we
don't do so by an excessive hero worship.''}

\medskip

And such a position is understandable, because Bell really did not proceed by using 
the standard for modern probability theory formalism of the Kolmogorov probability 
space and he did not point explicitly to violation of Kolmogorovness by data collected
in experimental tests of this inequality. However, the impact of his work to the probabilistic 
foundations of quantum mechanics and  the foundations of theory of probability in general was 
really great! Without Bell's works \cite{B0}, \cite{B} the development of quantum probability would be concentrated 
on merely mathematical issues without the direct connection with experiment. The key point is that
in Bell's framework the assumption on the existence of a single probability distribution serving to four incompatible 
measurements  and the possibility of its experimental testing enlightened (at least for experts in foundations of quantum 
mechanics and probability theory) the role of (non-)Kolmogorovness. Of course, the main problem was (as I pointed out in few my recent 
publications, e.g., \cite{arXiv0}) that Bell as the majority of physicists did not get a proper education in probability theory. Therefore
it was difficult for him to formulate  explicitly the thesis about the non-Kolmogorovian structure of his argument.  However, he 
prepared the soil and the harvest is really good, see, e.g., 
\cite{KB1}, \cite{book2} for reviews. Therefore I compared Bell with Copernicus, although I agree that 
comparisons with other heroes of science might be more appropriate.

Finally, I remark that in general Bell's works played the crucial role in stimulation 
of the presently growing interest to quantum foundations, e.g., \cite{T163}. Older experts in quantum foundations told me that in 1960th 
quantum foundations were not considered as a serious field of research in physics; they were treated merely as philosophical games.
As Alain Aspect told me, at his PhD defense  he was asked whether he had some job opportunities, and the members of jury were happy that 
he already got the offer for the permanent position. It was completely clear that with such a topic as the (first!) experimental violation 
of Bell's inequality he would never find job.  

\section{Conclusion}
\setcounter{equation}{0}

The analogy between the evolutions of geometry and probability and their impacts to physics support the scientific methodology based on construction 
of mathematical models of physical phenomena. By this methodology the development of physics  can be treated 
as transition from one mathematical model to another. We compare the contribution of Bell to probability with the 
contribution of Lobachevsky 
to geometry. Violations of Bell type inequalities imply rejection of the classical model of probability (Kolmogorov, 1933); quantum probability based 
on Born's rule is an example of non-Kolmogorovian probability. We propose to use this purely probabilistic interpretation  of 
violations of Bell type inequalities. Such an interpretation is very natural from the viewpoint of analogy with geometry -- the use 
of non-Euclidean models plays the fundamental role in special and general relativity. We also discuss the logical aspects of 
``Bell's revolution'' in foundations of probability as leading to rejection of Boolean logic and the applications of quantum logic.
Again by exploring the analogy between geometry and probability we come to the problem of embedding of a non-Kolmogorov probability (e.g., 
quantum probability) into the Kolmogorov model. We illustrate this problem by the CHSH-test (and probabilities generated by it)
as an example. We show that it is possible 
 to embed the CHSH-probabilities and correlations (which have the intrinsic non-Kolmogorovian structure) into  the Kolmogorov model.
 Our embedding is based on accounting randomness of selection of the experimental settings in the CHSH-test. In such an approach
 the quantum correlations are interpreted as the classical conditional correlations with respect to selection of the fixed experimental settings.
 Even in the classical (Kolmogorovian) probability approach such  conditional correlations can violate the CHSH-inequality. Thus
 we explain the violation of the CHSH-inequality by quantum correlations without appealing to nonlocality or rejecting realism.
 
As in geometry, the same quantum probabilistic structure can be embedded into the Kolmogorov model in various ways. In this paper we discussed 
only one possibility based on the treatment of quantum probabilities for the CHSH-experiment as conditional probabilities. And we used a special 
sort of conditioning: on the selection of the angles of PBSs. Other sorts of conditioning can serve for the same purpose. For example, in \cite{7X}
quantum correlations violating the CHSH-inequality were obtained as the result of conditioning on the simultaneous exceeding of the detection 
threshold by the two (of the four) components of a classical random field passing two (two channel) PBSs.   Conditioning on the simultaneous 
(more generally coincidence window) detection was used in \cite{D67}--\cite{D6},  but for particle-like classical systems (see \cite{THIS} for
 a review on the computer simulation approach to violations of Bell's inequality; see also \cite{Accardi} for another approach to computer simulation 
for violation of Bell's inequality).   

We remark that in this paper the present experimental status of violations of Bell's inequality was not discussed at all. The main reason for 
this is that our argument does not depend on complete justification of experimental violations of Bell's inequality, i.e., on closing of various loophole
(and first of all the locality and detection efficiency loopholes).

We point out that in  \cite{KH6}, \cite{KH7} Khrennikov and Volovich formulated an analog of the 
Heisenberg uncertainty principle for these loopholes (or more precisely the Bohr complementarity principle): they cannot be closed jointly in a single 
experimental test. (Here  the ``detection efficiency'' is treated as the efiiciency of the complete experimental scheme.) The argument used by the 
authors of \cite{KH6}, \cite{KH7} were of the heuristic nature. The recent studies on the asymptotic of the space-time dependence of the EPR-Bohm-Bell
correlations \cite{B1}--\cite{B3} support at least implicitly the Khrennikov-Volovich uncertainty principle for Bell's test. However, as was emphasized, from the viewpoint of the 
probabilistic opposition  all these problems with closing experimental loopholes are not important. Our position would not change even if the final loophole
free experimental test demonstrating violation of Bell's inequality were performed.     

Bell's inequality is just one of various tests of non-Kolmogorovness of quantum probabilities (if they are not treated as conditional probabilities).
For example, R. Feynman considered the two slit experiment as a test of nonclassicality of quantum probability \cite{Feynman}. He did not know
about the modern mathematical model of probability, Kolmogorov, 1933. Therefore he spoke about Laplacian classical probability. Feynman's argument 
in the modern probabilistic fashion was presented, e.g., in \cite{T1} -- \cite{T3}, \cite{INT}.

We recall that violations of Bell's inequality can be modeled not only with the aid of quantum probability, but even 
with the aid of other nonclassical probabilistic models.   For example, negative probabilities were widely explored, see \cite{INT} for 
a detailed review, see also \cite{N1}  for a recent study. In \cite{KH1}, \cite{KH3} I used so called $p$-adic probabilities  \cite{INT}, \cite{pp1}, \cite{pp2}. (In principle, such models are not less exotic than the quantum probabilistic model based
on representations of probabilities with the aid of Born's rule. The latter became commonly accepted through extensive applications.) Roughly speaking 
this is merely the matter of test whether to explore the complex probability amplitudes and Born's rule, or signed, or even $p$-adic valued 
 probabilities for the mathematical representations of quantum 
correlations. However, it seems that from the formal representational viewpoint the complex Hilbert space 
representation is the most convenient, in particular because it is linear.   

One can also proceed another way around: to represent the classical probabilities in the quantum-like manner, with the aid of complex probability 
amplitudes or in the abstract framework with the aid of vectors from complex Hilbert space. This is so-called {\it inverse Born's problem.} It was 
studied in very detail in \cite{book2}, see also \cite{Lin}, \cite{BAS}.

\section*{Appendix: Representation of events by subspaces, quantum logic}
\setcounter{equation}{0}

Bell's ``theorem'' is a consequences of the mathematical structure of QM.  While classical probability theory  
is based on the set-theoretical description, QM is founded on the premise
that events are associated with subspaces (or orthogonal projectors on these subspaces) of a vector space, complex Hilbert space. 
The adoption of subspaces as the basis  for predicting events also entails a new logic,  the logic of subspaces (projectors) 
which relaxes some of the axioms of classical Boolean logic (e.g., commutativity and distributivity). 

First time this viewpoint that QM is based on a new type of logic, {\it quantum logic}, was expressed in the book of von Neumann \cite{VN}, where 
he treated projectors corresponding to the eigenvalues of quantum observables (represented by Hermitian operators) as {\it propositions}, see also 
\cite{VBR}. The explicit 
formulation of logic of QM as a special quantum logic is based on the lattice  of all orthogonal projectors. 
For reader's convenience, below we present 
the mathematical structure of quantum logic. 

\subsubsection{Logical operations on for projectors}

For an orthogonal projector $P,$ 
we set $H_P= P(H),$ its image, and vice versa, for subspace $L$ of $H,$
the corresponding orthogonal projector is denoted by the symbol $P_L.$ 
 
The set of orthogonal projectors is a {\it lattice} with the order structure:
$P\leq Q$ iff $H_P \subset H_Q$ or equivalently, for any $\psi \in H, \; 
\langle \psi \vert P \psi  \rangle \leq \langle \psi \vert Q\psi  \rangle.$ 

We recall that the lattice of projectors is endowed with operations ``and'' ($\wedge$) and ``or'' ($\vee$). 
For two projectors $P_1, P_2,$ the projector $R=P_1 \wedge P_2$ is defined as the projector onto the subspace
$H_R= H_{P_1} \cap H_{P_2}$ and the projector $S=P_1 \vee P_2$ is defined as the projector onto the subspace
$H_R$ defined as the minimal linear subspace containing the set-theoretic union 
$H_{P_1} \cup H_{P_2}$ of subspaces $H_{P_1}, H_{P_2}:$ this is the space of all linear combinations of vectors
belonging these subspaces. The operation of negation is defined as the orthogonal complement.

In the language of subspaces the operation ``and'' coincides with the usual set-theoretic intersection, but the operations ``or'' and ``not'' 
are  nontrivial deformations of the corresponding set-theoretic operations. It is natural to expect that such deformations can induce 
deviations from classical Boolean logic. 

Consider the following simple example. Let $H$ be two dimensional Hilbert space with the orthonormal 
basis $(e_1,e_2)$ and let $v=(e_1+e_2)/\sqrt{2}.$ Then $P_v \wedge P_{e_1} =0$ and $ P_v \wedge P_{e_2}=0,$ 
but $ P_v \wedge (P_{e_1}\vee P_{e_2})=P_v.$ Hence, for quantum events,  in general the distributivity law is violated: 
\begin{equation}
\label{DIST}
P\wedge (P_{1}\vee P_{2}) \not= 
(P\wedge P_{1})\vee (P\wedge P_{2})
\end{equation}
As can be seen from our example, even mutual orthogonality of the events $P_1$ and $P_2$ does not help to save the Boolean laws.

We remark that for commuting projectors quantum logical operations have the Boolean structure. Thus noncommutativity can be considered 
as the algebraic representation of nonclassicality of quantum logic.   In particular, for a single observable with purely discrete spectrum
$A= \sum_\lambda \lambda P_\lambda,$ the projectors corresponding to different eigenvalues are orthogonal and, hence, they commutate. Therefore
deviations from classical logic and probability can be found only through analysis of results of a few incompatible measurements. 

A the first sight the representation of events by projectors/linear subspaces might look as exotic. However, this is simply a prejudice of 
the common use of the set-theoretic representation of events in the modern classical probability theory. The tradition to represent 
events by subsets was firmly established by A. N. Kolmogorov \cite{K}  only in 1933. We remark that before him the basic classical probabilistic models
were not of the set-theoretic nature. For example, the main competitor of the Kolmogorov model, the von Mises frequency model \cite{M1}, was based 
on the notion of a collective, see \cite{INT} for formulation of QM on the basis of the von Mises model.

\end{document}